\documentclass[12pt,a4paper,reqno,tbtags,oneside]{amsproc}
\usepackage{amsmath}
\usepackage{color}
\usepackage[colorlinks]{hyperref}

\email{m.v.pavlov@lboro.ac.uk} \subjclass{MSC subject
classification }
\input{tcilatex}

\begin{document}
\title{The Boussinesq equation and Miura type transformations}
\date{\today }
\author{Maxim Pavlov}
\address{Institute of Mathematics, Academia Sinica, Taipei 11529, Taiwan}

\begin{abstract}
Several Miura type transformations for the Boussinesq equation are found and
the corresponding integrable systems constructed.
\end{abstract}

\maketitle
\tableofcontents

\section{Introduction}

In \cite{Maks} integrable equations that admit a scalar spectral problem
were considered and an algorithm for constructing the Miura type
transformations for these equations was described. The KdV equation was
considered as an example. Another example, the Kaup-Boussinesq equation, was
studied in \cite{Maks1}. The latter equation is related with the NLS
equation by a pair of invertible differential first order substitutions. In
fact, there exist more powerful methods that allow to describe modified
equations, that is, the equations related with the initial ones by
non-invertible differential substitutions. These substitutions are called
the Miura type transformations. The approach based on constructing the
dressing sequences of discrete symmetries was elaborated in the paper \cite%
{Borisov}. The KdV equation and the Bonnet equation served as examples in
that paper (the Bonnet equation is also known as the sine-Gordon equation in
the mathematical physics literature). The Kaup-Boussinesq equation was
considered in the paper \cite{Borisov1}. This method allows to construct
\textit{multi-parametric} integrable systems and does not depend on the form
of the spectral problem, which can be either scalar or matrix. Therefore
this method differs from the one proposed in \cite{Maks}. One may think that
the first approach was proposed in order to replicate integrable equations.
This is not so. Conversely, new integrable equations were obtained as an
auxiliary result while decreasing the orders of the Hamiltonian operators.
Indeed, the reduction of the Hamiltonian operators to the canonical form
\textquotedblleft $d/dx$\textquotedblright\ was the aim of that approach,
which can be treated as a generalization of the Darboux theorem on reduction
of a Hamiltonian operator to the constant operator for the
infinite-dimensional case. Also, we emphasize that the method described in
\cite{Maks} is extremely simple from the technical viewpoint. Still, this
procedure was not applied for the Boussinesq equation which is at the next
level of complexity in comparison with the KdV equation. The present paper
is devoted to solution of this problem. Since the Boussinesq equation is a
two-component system of equations, then new field variables, which are used
in construction of the modified equations, are not uniquely defined.
Moreover, the third order spectral problem admits two sets of non-invertible
differential substitutions, unlike the case of the KdV or the
Kaup-Boussinesq equations, which are related with the second order spectral
problem. We also see that any non-invertible differential substitution
decreases the order of the Hamiltonian operator such that the canonical form
\textquotedblleft $d/dx$\textquotedblright\ for the KdV and the
Kaup-Boussinesq equation is achieved. Now consider the Boussinesq equation.
Then only one of two sets of the Miura type transformations is related with
the Hamiltonian structures. Namely, this is the set defined by
transformations that are \textit{quadratic} in the field variables; another
set is \textit{cubic}. This situation illustrates a general concept: any
scalar spectral problem of order $N$ admits a unique sequence of
differential substitutions related with the Hamiltonian operators. Indeed,
the field variables, which appear in powers not greater than two in the
corresponding set, are conservation law densities for the modified
equations, simultaneously. By (\cite{Maks}, \cite{Maks1}, \cite{Maks+Tsar}),
if the number of the field variables used in the quadratic Miura type
transformation equals $M+1$, where $M$ -- is the number of equations in the
corresponding integrable system, then there are $M$ variables that
annihilate the Poisson bracket which is related with the zero curvature
metric (in case the coordinates are flat) or the constant curvature metric.
If the number of summands in such quadratic transformation exceeds $M+1$,
then the corresponding Hamiltonian structure is related with the nonlocal
Ferapontov brackets (see \cite{Fer}, \cite{Malt}). In this paper, we do not
concentrate on the reduction of the Hamiltonian operators to the canonical
form \textquotedblleft $d/dx$\textquotedblright , Indeed, all our attention
is drawn to the description of all possible differential substitutions
(similar to ones obtained for the KdV equation) in view of the specific
features discussed above (e.g., the presence of two components in the system
at hand and the third order scalar spectral problem).

The method proposed in \cite{Maks} is based on the following reasoning.
Consider the integrable system defined by an $\hat{L}\hat{A}$ pair such that
the scalar differential operator $\hat{L}$ of order $N$ is polynomial in the
spectral parameter $\lambda $ (Note that only four examples of such
operators are known for the simplest case $N=3$. They are: systems that can
be reduced to the Boussinesq equation by using invertible differential
substitutions, to the long-short resonance system, to the two-component NLS,
and, probably, to the 3-wave system, see \cite{Fordy}). Consider the
equation $\hat{L}\psi =0$ and make the substitution%
\begin{equation}
\psi =\exp [\int rdx].  \label{1}
\end{equation}%
Hence we obtain a nonlinear differential equation, which is a generalization
of the Riccati equation for $N>2$. Now, decompose $r$ to the Laurent series
in $\lambda $ in vicinity of the infinity. We get the infinite set of
differential polynomials in the field variable. Each polynomial is a
conservation law density. Our main idea is that one must consider the
decomposition of $r$ to the Taylor series in $\lambda $ in vicinity of zero!
The initial coefficients of the series
\begin{equation*}
r=a+\lambda b+\lambda ^{2}c+...
\end{equation*}%
are the new field variables. Then we represent the modified integrable
systems by using these variables, while the relations between the old and
new field variables are precisely the Miura type transformations.

\section{The \textit{cubic} Miura type transformations}

Represent the Boussinesq equation%
\begin{equation*}
u_{tt}=\partial _{x}^{2}[-\frac{1}{3}u_{xx}+\frac{2}{3}u^{2}],
\end{equation*}%
as the system of two evolution equations
\begin{equation}
u_{t}=\partial _{x}\eta \text{, \ \ \ \ }\eta _{t}=\partial _{x}[-\frac{1}{3}%
u_{xx}+\frac{2}{3}u^{2}].  \label{3}
\end{equation}%
This equation is the compatibility condition for two linear differential
equations%
\begin{equation}
\psi _{xxx}=u\psi _{x}+(\lambda ^{3}+\frac{1}{2}\eta +\frac{1}{2}u_{x})\psi
\text{, \ \ \ \ \ }\psi _{t}=\psi _{xx}-\frac{2}{3}u\psi .  \label{4}
\end{equation}%
By using substitution (\ref{1}) we rewrite system (\ref{4}) in the form
\begin{equation}
\begin{array}{c}
r_{xx}+3rr_{x}+r^{3}=ru+\frac{1}{2}\eta +\frac{1}{2}u_{x}+\lambda ^{3}, \\
\\
r_{t}=\partial _{x}[r_{x}+r^{2}-\frac{2}{3}u],%
\end{array}
\label{6}
\end{equation}%
Here the first equation is the generating function of conservation law
densities (as $\lambda \rightarrow \infty $) and the second equation is the
generating function of conservation laws. Now substitute the Taylor series%
\begin{equation}
r=a+\lambda ^{3}b+\lambda ^{6}c+...  \label{2}
\end{equation}%
\ in the first equation among these two. Hence we obtain the Miura type
transformations%
\begin{equation}
\begin{array}{c}
a_{xx}+3aa_{x}+a^{3}=ua+\frac{1}{2}\eta +\frac{1}{2}u_{x}, \\
\\
b_{xx}+3ab_{x}+3ba_{x}+3a^{2}b=ub+1, \\
\\
c_{xx}+3ac_{x}+3bb_{x}+3ca_{x}+3a^{2}c+3ab^{2}=uc,...%
\end{array}
\label{7}
\end{equation}%
Now substitute the Taylor series given by (\ref{2}), in the second
equation. Then we obtain the corresponding pseudononlocal
conservation laws, that is, the conservation laws such that the
densities and the fluxes are not expressed in terms of the field
variables $u$, $\eta $ and \textit{finite}
number of their derivatives: we have%
\begin{equation}
\begin{array}{c}
a_{t}=\partial _{x}[a_{x}+a^{2}-\frac{2}{3}u], \\
\\
b_{t}=\partial _{x}[b_{x}+2ab], \\
\\
c_{t}=\partial _{x}[c_{x}+2ac+b^{2}],...%
\end{array}
\label{8}
\end{equation}%
for the Boussinesq equation. Now express $\eta $ from the first equation in (%
\ref{7}). We get%
\begin{equation*}
\eta =2a_{xx}+6aa_{x}+2a^{3}-2au-u_{x},
\end{equation*}%
Hence we obtain the modified Boussinesq equation (MB)%
\begin{eqnarray}
a_{t} &=&\partial _{x}[a_{x}+a^{2}-\frac{2}{3}u],  \label{9} \\
u_{t} &=&\partial _{x}[2a_{xx}+6aa_{x}+2a^{3}-2au-u_{x}].  \notag
\end{eqnarray}%
Next, express $u$ from the second equation in (\ref{7}). We get%
\begin{equation}
u=3(a_{x}+a^{2})+\frac{b_{xx}+3ab_{x}-1}{b},  \label{16}
\end{equation}%
Hence we obtain the twice modified Boussinesq equation (TwMB)
\begin{eqnarray}
b_{t} &=&\partial _{x}[b_{x}+2ab],  \label{10} \\
a_{t} &=&\partial _{x}[-a_{x}-a^{2}-\frac{2}{3b}(b_{xx}+3ab_{x}-1)].  \notag
\end{eqnarray}%
Now, express $a$ from the third equation in (\ref{7})
\begin{equation*}
a=\frac{c-(cb_{x}-bc_{x}-b^{3})_{x}}{3(cb_{x}-bc_{x}-b^{3})}.
\end{equation*}%
Naturally, we get the thrice modified Boussinesq equation (ThrMB)
\begin{eqnarray}
c_{t} &=&\partial _{x}[c_{x}+b^{2}+2c\frac{c+3b^{2}b_{x}+bc_{xx}-cb_{xx}}{%
3(cb_{x}-bc_{x}-b^{3})}],  \label{11} \\
b_{t} &=&\partial _{x}[b_{x}+2b\frac{c+3b^{2}b_{x}+bc_{xx}-cb_{xx}}{%
3(cb_{x}-bc_{x}-b^{3})}].  \notag
\end{eqnarray}%
From general theory of linear ordinary differential equations it is well
known that the Wronskian of three linearly independent solutions ($\psi $, $%
\psi ^{-}$, $\psi ^{+}$) to the first equation in (\ref{4}) is equal to a
constant:
\begin{equation*}
(s_{xx}-us)\psi -s_{x}\psi _{x}+s\psi _{xx}=\varepsilon .
\end{equation*}%
Here $s=\psi ^{-}\psi _{x}^{+}-\psi ^{+}\psi _{x}^{-}$ is a solution of a
conjugate equation (see first equation in (\ref{4}))%
\begin{equation*}
s_{xxx}=us_{x}-(\lambda ^{3}+\frac{1}{2}\eta -\frac{1}{2}u_{x})s.
\end{equation*}

\textbf{Theorem 1}: \textit{The function}\textbf{\ }$\varphi =s\psi $
\textit{is a solution of the ordinary differential equation}%
\begin{equation}
\varphi _{xx}-3r\varphi _{x}+(3r^{2}-u)\varphi =\varepsilon  \label{12}
\end{equation}%
\textit{Also, this function is the generating function of conservation law
densities for the Boussinesq equation}%
\begin{equation}
\varphi _{t}=\partial _{x}[2r\varphi -\varphi _{x}].  \label{13}
\end{equation}

The proof follows from (\ref{4}) by a straightforward calculation.

\textbf{Remark 1}: The function $\varphi $ is not a new generating function
for conservation law densities. Indeed,%
\begin{equation*}
\frac{\delta R}{\delta \eta }=\frac{1}{2\varepsilon }\varphi \text{,}
\end{equation*}%
where $R=\int rdx$. The latter equality means that the Euler derivative $%
\delta /\delta \eta $ acts as the shift operator of the space of
conservation law densities%
\begin{equation*}
\frac{\delta H_{k+1}}{\delta \eta }=h_{k},
\end{equation*}%
where $H_{k}=\int h_{k}dx$. Therefore, the coefficients of the Laurent
series in $\lambda $ around infinity for the function $\varphi $ (see (\ref%
{12}) and (\ref{13})) differ from the coefficients of the Laurent series for
the function $r$ by some constant factors only (see also (\ref{6})).

Recall that for the KdV equation it was sufficient to express the old field
variable by the new one. Next, consider the Kaup-Boussinesq equation, which
is a two-component system. Then this procedure is already not sufficient for
the modified system to be uniquely defined \cite{Borisov}. Indeed, there are
two possible variants in this case (see (\ref{12})). The first case is%
\begin{equation*}
u=3r^{2}+\frac{\varphi _{xx}}{\varphi }-3r\frac{\varphi _{x}}{\varphi }-%
\frac{\varepsilon }{\varphi },
\end{equation*}%
The first modified Boussinesq equation (MB$_{1}$)%
\begin{equation}
r_{t}=\partial _{x}[r_{x}-r^{2}+2r\frac{\varphi _{x}}{\varphi }-\frac{%
2\varphi _{xx}}{3\varphi }+\frac{2\varepsilon }{3\varphi }]\text{, \ \ \ \ \
\ }\varphi _{t}=\partial _{x}[2r\varphi -\varphi _{x}]  \label{0}
\end{equation}%
is of orders 3 and 2 with respect to the derivatives. Still, the second
modified Boussinesq equation (MB$_{2}$)%
\begin{equation*}
u_{t}=\partial _{x}[2r_{xx}+6rr_{x}+2r^{3}-2ru-u_{x}]\text{, \ \ \ \ }%
\varphi _{t}=\partial _{x}[2r\varphi -\varphi _{x}],
\end{equation*}%
where%
\begin{equation*}
r=\frac{\varphi _{x}}{2\varphi }\pm \sqrt{-\frac{\varphi _{xx}}{3\varphi }+%
\frac{\varphi _{x}^{2}}{4\varphi ^{2}}+\frac{\varepsilon }{3\varphi }+\frac{1%
}{3}u},
\end{equation*}%
is of orders 5 and 3, respectively. The leading order with respect to
derivatives is preserved by the Miura type transformations for the scalar
KdV type equations. We see that this is not true even in the two-component
case. The explicit formulas are huge and therefore omitted.

\textbf{Remark 2}: Invertible differential substitutions $a=p-b_{x}/(2b)$
for TwMB (\ref{10}) and $q=r-\varphi _{x}/(2\varphi )$ for the MB$_{1}$ (\ref%
{0}) yield the same modified system
\begin{equation*}
\begin{array}{c}
b_{t}=\partial _{x}(2bp), \\
\\
p_{t}=\partial _{x}[-p^{2}+\frac{2}{3b}+\frac{b_{x}^{2}}{4b^{2}}-\frac{b_{xx}%
}{6b}],%
\end{array}%
\text{ \ \ \ \ \ \ \ \ \ \ \ \ \ \ \ \ \ \ \ \ \ \ \ \ }%
\begin{array}{c}
\varphi _{t}=\partial _{x}(2\varphi q), \\
\\
q_{t}=\partial _{x}[-q^{2}+\frac{2\varepsilon }{3\varphi }+\frac{\varphi
_{x}^{2}}{4\varphi ^{2}}-\frac{\varphi _{xx}}{6\varphi }].%
\end{array}%
\end{equation*}

\section{The \textit{quadratic} Miura type transformations}

In the Introduction we noted that only quadratic Miura type transformations
are related with the Hamiltonian structures. Now, consider the factorization
of the first equation in scalar spectral problem (\ref{4}). We have%
\begin{equation*}
(\partial _{x}+a+\bar{a})(\partial _{x}-\bar{a})(\partial _{x}-a)\psi
=\lambda ^{3}\psi
\end{equation*}%
This factorization supplies the well-known Miura type transformation (see
\cite{Fordy1})%
\begin{equation}
u=2a_{x}+\bar{a}_{x}+a^{2}+a\bar{a}+\bar{a}^{2}.  \label{18}
\end{equation}%
Therefore, the third modified Boussinesq equation (MB$_{3}$)%
\begin{eqnarray}
a_{t} &=&\frac{1}{3}\partial _{x}[a^{2}-2a\bar{a}-2\bar{a}^{2}-a_{x}-2\bar{a}%
_{x}],  \label{14} \\
\bar{a}_{t} &=&\frac{1}{3}\partial _{x}[-2a^{2}-2a\bar{a}+\bar{a}^{2}+2a_{x}+%
\bar{a}_{x}]  \notag
\end{eqnarray}%
is the compatibility condition for%
\begin{eqnarray*}
\left(
\begin{array}{c}
\psi \\
\psi _{1} \\
\psi _{2}%
\end{array}%
\right) _{x} &=&\left(
\begin{array}{ccc}
a & \lambda & 0 \\
0 & \bar{a} & \lambda \\
\lambda & 0 & -a-\bar{a}%
\end{array}%
\right) \left(
\begin{array}{c}
\psi \\
\psi _{1} \\
\psi _{2}%
\end{array}%
\right) , \\
&& \\
\left(
\begin{array}{c}
\psi \\
\psi _{1} \\
\psi _{2}%
\end{array}%
\right) _{t} &=&\left(
\begin{array}{ccc}
a_{x}-\frac{2}{3}u+a^{2} & \lambda (a+\bar{a}) & \lambda ^{2} \\
\lambda ^{2} & \frac{1}{3}u-a^{2}-a\bar{a} & -\lambda a \\
-\lambda \bar{a} & \lambda ^{2} & \frac{1}{3}u+a\bar{a}-a_{x}%
\end{array}%
\right) \left(
\begin{array}{c}
\psi \\
\psi _{1} \\
\psi _{2}%
\end{array}%
\right) .
\end{eqnarray*}%
Two local Hamiltonian structures for the Boussinesq equation are well known.
The first structure is of the canonical form (see for instance, \cite{Olver}%
): we have
\begin{equation}
u_{t}=\partial _{x}\frac{\delta H_{4}}{\delta \eta }\text{, \ \ \ \ \ }\eta
_{t}=\partial _{x}\frac{\delta H_{4}}{\delta u},  \label{17}
\end{equation}%
such that the Hamiltonian is $H_{4}=\int [\frac{1}{2}\eta ^{2}+\frac{1}{6}%
u_{x}^{2}+\frac{2}{9}u^{3}]dx$, the momentum is $H_{3}=\int u\eta dx$, and
two annihilators (Casimirs) are $H_{2}=\int \eta dx$, \ $H_{1}=\int udx$.
The other Hamiltonian structure is%
\begin{eqnarray*}
u_{t} &=&\partial _{x}[\frac{3}{2}\eta \frac{\delta H}{\delta \eta }%
+(-\partial _{x}^{2}+u)\frac{\delta H}{\delta u}]-\frac{1}{2}(\frac{\delta H%
}{\delta \eta }\eta _{x}+\frac{\delta H}{\delta u}u_{x}) \\
\eta _{t} &=&\partial _{x}(\frac{1}{3}[\partial _{x}^{4}-5u\partial _{x}^{2}-%
\frac{5}{2}u_{x}\partial _{x}+2(-u_{xx}+2u^{2})]\frac{\delta H}{\delta \eta }%
+ \\
&&+\frac{3}{2}\eta \frac{\delta H}{\delta u})-[\frac{\delta H}{\delta u}%
\partial _{x}\frac{\delta H_{4}}{\delta \eta }+\frac{\delta H}{\delta \eta }%
\partial _{x}\frac{\delta H_{4}}{\delta u}].
\end{eqnarray*}%
Consider the variables ($a$, $\bar{a}$). Then the second structure acquires
the canonical form
\begin{equation}
a_{t}=\frac{1}{3}\partial _{x}[-2\frac{\delta H_{2}}{\delta a}+\frac{\delta
H_{2}}{\delta \bar{a}}]\text{, \ \ \ \ \ }\bar{a}_{t}=\frac{1}{3}\partial
_{x}[\frac{\delta H_{2}}{\delta a}-2\frac{\delta H_{2}}{\delta \bar{a}}],
\label{19}
\end{equation}%
such that the Hamiltonian is $H_{2}=-\frac{1}{2}\int \eta dx=\int a[\bar{a}%
(a+\bar{a})+\bar{a}_{x}]dx$, the momentum is $H_{1}=-\int udx=-\int [a^{2}+a%
\bar{a}+\bar{a}^{2}]dx$, and two annihilators are $H_{-1}=\int adx$, \ $\bar{%
H}_{-1}=\int \bar{a}dx$. We conclude that MB$_{3}$ admits local Hamiltonian
structure (\ref{19}).

Finally, we describe another modified Boussinesq equation ThrMB$_{2}$%
\begin{equation*}
b_{t}=\partial _{x}[b_{x}+2ab]\text{, \ \ \ \ \ \ \ }s_{t}=\partial
_{x}[s(s+2a)+(s+2a)_{x}],
\end{equation*}%
where $s=\bar{a}-a$. The Miura type transformation%
\begin{equation*}
a=\frac{-b_{xx}+(s^{2}+s_{x})b+1}{b_{x}-sb}
\end{equation*}%
is obtained by excluding $u$ in (\ref{16}) and (\ref{18}). Therefore, ThrMB$%
_{2}$ is of orders 3 and 4 with respect to the higher variables in its first
and second components, respectively.

\section{Concluding remarks}

In this paper, we have found all \textquotedblleft
obvious\textquotedblright\ Miura type transformations that preserve the
conservative form of the modified equations. Suppose the latter restriction
is omitted. Then the Muira type transformations are still few, although
their number increases (e.g., the KdV equation admits two transformations
that preserve the conservative form and one that spoils it). Moreover,
suppose that transformations of the independent variables ($x$, $t$) are
also allowed. Then we can continue the sequence of the modified equations,
see \cite{Maks}, \cite{SS}). Such procedure will be described elsewhere.
Also full set of multi-parametric modified Boussinesq equations by approach
given in \cite{Borisov} should be found.

In \cite{Shabat} classification of integrable systems%
\begin{equation*}
u_{t}=u_{xx}+F(u\text{, }w\text{, }u_{x}\text{, }w_{x})\text{, \ \ \ \ \ \ }%
-w_{t}=w_{xx}+G(u\text{, }w\text{, }u_{x}\text{, }w_{x})
\end{equation*}%
was presented. The Boussinesq equation belongs to this class (by virtue of
invertible differential substitution $\eta =\rho \pm iu_{x}/\sqrt{3}$). It
will be nice to prove that all above modified Boussinesq equations up to
module of invertible differential substitutions also belong to this class.

\section*{Acknowledgments}

I would like to thank Dr. Arthemy Kiselev who suggested me to
write this paper (it was a talk at International Conference on
"Integrable Systems: Solutions and Transformations" Guardamar
(Alicante, Spain) at 15-19 June 1998).

I would like to thank the Loughborough University, UK for their
financial support and hospitality when this work was made.

\end{document}